\documentclass[preprint]{revtex4}
\usepackage{bm}
 \usepackage{epsfig}
\input{epsf}
\begin{document}
\title{Electronic band structure engineering in InAs/InSbAs and InSb/InSbAs superlattice heterostructures}

\author{Atanu Patra$^{1}$, Monodeep Chakraborty$^{2}$, Anushree Roy$^{1}$}
 \affiliation{
  $^{1}$Department of Physics, Indian Institute of Technology Kharagpur, Kharagpur 721302, India\\
  $^{2}$Center for Theoretical Studies,\\ Indian Institute of Technology Kharagpur, Kharagpur 721302, India}


\begin{abstract}
We report a detailed \emph{ab initio} study of two superlattice
heterostructures, one component of which is a unit cell of CuPt
ordered InSb$_{0.5}$As$_{0.5}$. This alloy part of the
heterostructures is a topological semimetal.  The other component of
each system is a semiconductor, zincblende-InSb, and wurtzite-InAs.
Both heterostructures are semiconductors. Our theoretical analysis
predicts that the variation in the thickness of the InSb layer in
InSb/InSb$_{0.5}$As$_{0.5}$ heterostructure  renders altered band
gaps with different characteristics (\emph{i.e.} direct or
indirect). The study holds promise for fabricating heterostructures,
in which the modulation of the thickness of the layers changes the
number of carrier pockets in these systems.

\end{abstract}
\maketitle
\def\d{{\mathrm{d}}}

\section{Introduction}
Low band gap materials, like InSbAs alloys, find potential
applications in various infra-red device fabrications
\cite{Pitanti1,Hoglund1,Steveler1,Jiang1,Wu1}. In addition, due to
large lande-g factor, significant spin-orbit coupling strength and
small effective mass, these compounds are also used  in high-speed
\cite{Ashley,Sourribes1} and spin-related \cite{Nilsson} devices.
The promising characteristics of
InSb$_{x}$As$_{1-x}$/InSb$_{y}$As$_{1-y}$ heterostructures (HSs) in
various applications motivated us to investigate electronic band
structure of two superlattice HSs.  One component of both HSs
consists of one unit cell of InSb$_{0.5}$As$_{0.5}$ in CuPt
ordering. For one of the HSs, the other component is InSb in
zinc-blende (ZB) structure along [111] direction. For the second HS,
the other component is InAs in wurtzite (WZ) phase along [0001]
direction. We varied the thickness of InAs or InSb segments. Thus,
we have studied band structures of two
 sets of systems, (InSb)$_n$(ZB)/InSb$_{0.5}$As$_{0.5}$(CuPt) and
(InAs)$_m$(WZ)/InSb$_{0.5}$As$_{0.5}$(CuPt), named as H-I and H-II,
respectively. Here, $n$ and $m$ are the number of unit cells of InSb
and InAs  segments in H-I and H-II. The choice of two different
phases (i.e. ZB and WZ) of InSb and InAs parts originates from the
fact that while growing these HSs (eg., in HS nanowires) using MOCVD
or CBE techniques, InSb and InAs components are formed in these
phases \cite{Johansson,Ercolani1}. Moreover, these two different
semiconductor layers result in compressive and tensile strain at the
heterointerface with CuPt-InSb$_{0.5}$As$_{0.5}$.

In this article, we demonstrate that the band gap of the HSs can be
tuned by the choice of semiconductor segments and their thickness
(\emph{i.e.}, $m$ and $n$). Interestingly, for H-I with $n$=1 both
direct and indirect band gaps are of equal energy. In addition, the
calculated band structure opens a possibility of achieving different
number of carrier pockets in these HSs under perturbation.

\section{Methodology}

 We have carried
out first principles calculations with Wien2k, which is an
all-electron-full potential code \cite{Singh1,Blugel,Blaha1}. All
crystal structures, discussed in this article, were optimized
following Ref. \cite{Patra1}. To study the electronic band
structure, we have performed calculations using local density
approximation (LDA) and through modified Becke-Johnson exchange
potential (mBJLDA) \cite{Tran1}. The spherical harmonic function
inside the muffin-tin spheres was limited by \emph{l}$_{max}$ =
$12$, where the muffin-tin radii for In, As and Sb were fixed at
$1.9$, $2.1$ and $2.15$ a.u., respectively. In interstitial regions
the charge density and potential were defined through
\emph{G}$_{max}$ at $14$ Bohr$^{-1}$. The tetrahedron method was
employed for Brillouin zone (BZ) integrations within
self-consistency cycles \cite{Blochl1}. The basis set convergence
parameter ($R_{MT}^{min}K_{max}$)
 was set to $8$ for all calculations.

The $k$-mesh for the band structure of CuPt ordered
InSb$_{0.5}$As$_{0.5}$ was $14\times14\times14$. The calculations
for HSs were carried out with a $k$-mesh of $20\times20\times1$.
 To obtain high accuracy in our calculations the  effect of spin-orbit coupling (SOC) was implemented through a second variational procedure,
where states up to 9 Ry above Fermi energy ($E_F$) were included in
the basis expansion and the relativistic $p_{1/2}$ corrections were
incorporated for the higher  lying \emph{p} orbitals.

\section{Results and Discussion}
\begin{figure}[!t]
 \centering
\includegraphics[scale=0.4]{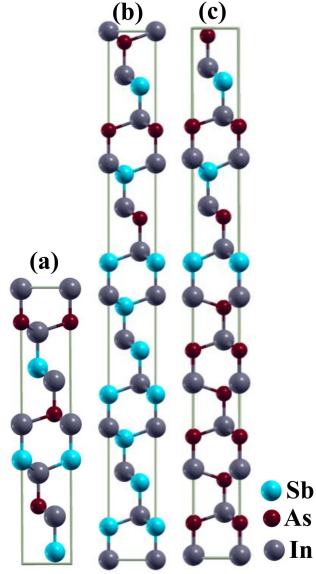}
\caption{One unit cell of (a) CuPt ordered InSb$_{0.5}$As$_{0.5}$,
(b) H-I (\emph{n}=2) and (c) H-II (\emph{m}=3).} \label{structure}
\end{figure}
\subsection{Structural properties}

 The atomic arrangement in a unit cell of CuPt ordered
InSb$_{0.5}$As$_{0.5}$ is shown in Fig. \ref{structure} (a). The
atomic arrangement in optimized HS unit cells of H-I with \emph{n}=2
and H-II with \emph{m}=3 are shown in Fig. \ref{structure} (b) and
(c). The optimized structures of other HSs, i.e. with $n$=1,3 (in
H-I) and $m$=1,2 (in H-II) are also obtained. A unit cell of CuPt
ordered InSb$_{0.5}$As$_{0.5}$ has 12 layers of atoms. A unit cell
of a ZB-InSb [111]  has six layers of atom; whereas, the same for a
WZ structure in [0001] direction has four layers of atomic
arrangement. Thus, the number of layers in H-I are 18 (\emph{n}=1),
24 (\emph{n}=2) and 30 (\emph{n}=3) whereas in H-II, the values are
16 (\emph{m}=1), 20 (\emph{m}=2) and 24 (\emph{m}=3). The optimized
lattice constant for CuPt ordered InSb$_{0.5}$As$_{0.5}$ are $ a =b=
4.4586$ \AA\ and $ c = 21.8427$ {\AA} with the  space group $R3m$
(no. 160). Both H-I and H-II take a trigonal structure with the
space group $P3m1$ (no. 156). The lattice parameters of H-I
(\emph{n}=1--3) and H-II (\emph{m}=1--3) are listed in
Table-\ref{l.c.}.
\begin{table}[ht]
\setlength\belowcaptionskip{5pt}\caption{The optimized lattice constants of H-I and H-II with different \emph{n} and \emph{m}.} 
\centering %
\begin{tabular}{c c c c c c c c c c c c c c c c c c c}
\hline
HS & & \emph{a} && \emph{c} & \\
  & & in {\AA} &&  in {\AA} & \\
\hline\hline
   & \emph{n}     \\
&1 &4.6077&&33.8598 &\\
\textbf{H-I} &2 &4.6154&&45.2216 &\\
&3 &4.6231&&56.6206 &\\
\hline
   & \emph{m}     \\
&1 &4.4935&&29.3514 &\\
\textbf{H-II} &2 &4.4636&&36.4449 &\\
&3 &4.3752&&43.9056 &\\
\hline
\hline %
\setlength\belowcaptionskip{5pt}
\end{tabular}
\label{l.c.}
\end{table}

\subsection{Electronic properties}

\begin{figure}[t]
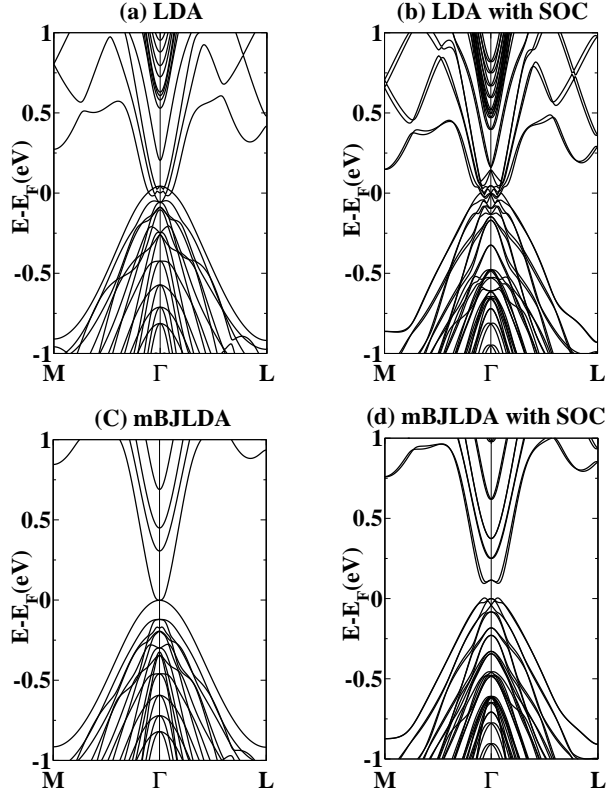

\vskip 0.3cm
\hspace{0.0in}\includegraphics[scale=0.18]{HI_2_LDA.eps}
\hspace{0.3in}\includegraphics[scale=0.18]{HI_2_LDA_SO.eps} \vskip
0.3cm \hspace{0.0in}\includegraphics[scale=0.18]{HI_2_MBJ.eps}
\hspace{0.3in}\includegraphics[scale=0.18]{HI_2_MBJ_SO.eps}
\caption{The band structure of H-I (\emph{n}=2) using different
procedures under (a) LDA, (b) LDA with SOC (c) mBJLDA (d) mBJLDA
with SOC schemes.} \label{band-H-I}
\end{figure}

 We have carried out systemic band
structure calculations on two HSs, H-I and H-II with
(\emph{n,m}=1--3), using LDA, LDA with SOC, mBJLDA and mBJLDA with
SOC schemes to demonsrate the role of the mBJLDA exchange potential
and SOC in the band structure. From the calculated band structure,
we plot the energy bands along $M$-$\Gamma$-$L$ \emph{k}-paths. Fig.
\ref{band-H-I} (a)--(d) present the same for  H-I (\emph{n}=2). The
standard LDA method yields this system as a gapless material (Fig.
\ref{band-H-I} (a)), However, the implementation of SOC interaction
over LDA does not open a gap in the system (Fig. \ref{band-H-I}
(b)). The mBJLDA exchange potential creates a  separation between
the bands, present near the E$_F$ (see Fig. \ref{band-H-I} (c)).
Nonetheless, the band structure still exhibits metallic behavior.
Incorporation of SOC over the mBJLDA calculation, as shown in Fig.
\ref{band-H-I} (d), shows a band gap. Similarly, for H-I with
$n$=1,3, the band gap could be obtained only using mBJLDA exchange
potential with SOC. Above mentioned other calculation schemes showed
the metallic behavior of the systems.

\begin{figure}[h!]
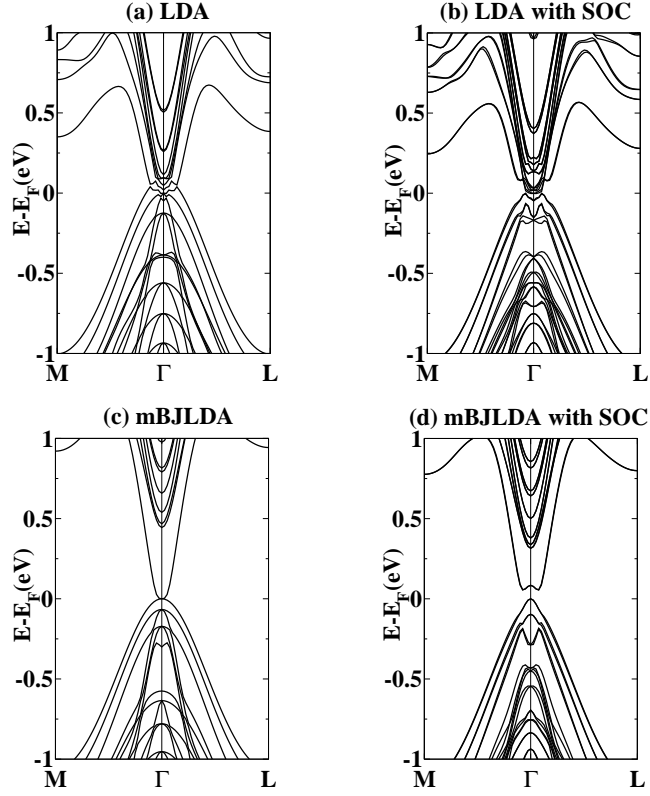

\vskip 0.3cm \hspace{0.0in}\includegraphics[scale=0.18]{fig2a.eps}
\hspace{0.5in}\includegraphics[scale=0.18]{fig2b.eps} \vskip 0.3cm
\hspace{0.0in}\includegraphics[scale=0.18]{fig2c.eps}
\hspace{0.5in}\includegraphics[scale=0.18]{fig2d.eps} \caption{The
band structure of H-II (\emph{m}=3) using different procedures under
(a) LDA, (b) LDA with SOC (c) mBJ and (d) mBJ with SOC schemes.}
\label{band-HII}
\end{figure}

We have performed the calculations on H-II, with $m$=1--3, following
the same four-steps procedure as used for H-I. H-II (with $m$=1--3)
exhibited the same trend, as observed in the case of H-I. Under LDA,
LDA with SOC and mBJLDA schemes, we obtained metallic behavior of
this system. Incorporation of SOC over mBJ corrected LDA calculation
opens a  gap at the $\Gamma$ point. In Fig. \ref{band-HII} (a)--(d)
we plot the energy bands of H-II with $m$=3, as obtained using above
the mentioned schemes.

\begin{figure}
\vskip 0.3cm \hspace{0.1in}\includegraphics[scale=0.6]{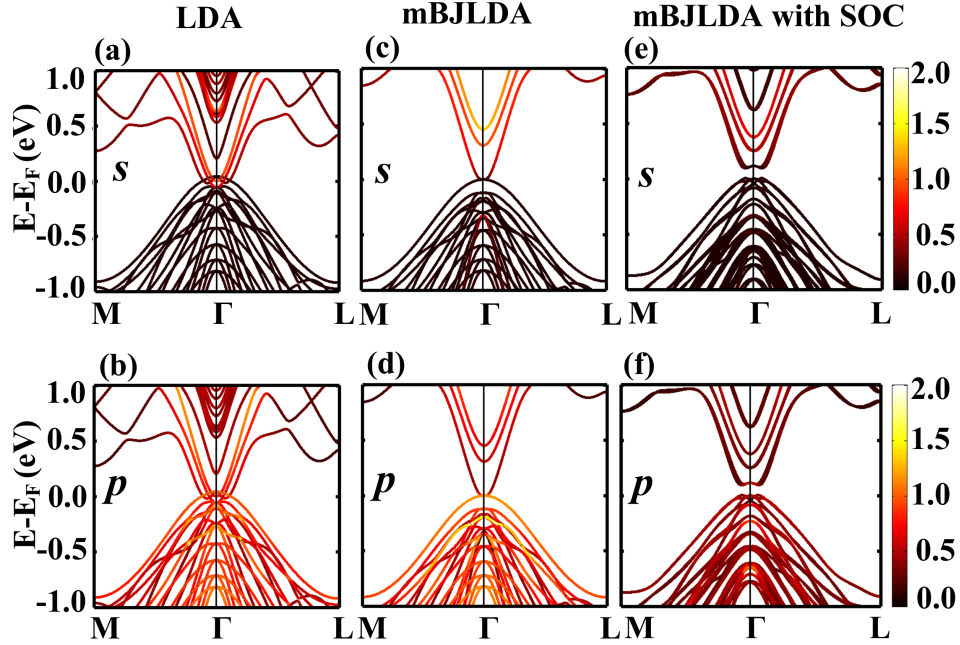}
\caption{Fat bands showing the contributions of (a)  $s$ and (b) $p$
orbitals in H-I (\emph{n}=2) with LDA.  The same of (c)  $s$ and (d)
 $p$ orbitals with mBJLDA+SOC scheme.}\label{fat-HI}\end{figure}

We note that the LDA underestimates the separation between different
bands to a great extent.  It is to be noted that for InSb based
materials Sb-\emph{p} states  dominate near $E_F$ \cite{Cardona1}.
Fat band analysis provides the contribution of atomic orbitals in
the band structure. Fat bands showing the contribution of $s$ and
$p$ orbitals in the band structure of H-I with \emph{n}=2 under the
LDA scheme are plotted in Fig. \ref{fat-HI} (a)--(b), mBJLDA in Fig.
\ref{fat-HI} (c)--(d) and under mBJLDA+SOC in Fig. \ref{fat-HI}
(e)--(f). While the upper three panels, (a), (c) and (e), reveal the
contribution of $s$ orbitals, the lower three panels (b), (d) and
(d), demonstrate the contribution of the $p$ orbitals. We find that
under LDA, In/As \emph{s} and Sb \emph{p}-orbitals mostly contribute
near $E_F$. The spin-orbit interaction becomes significant for high
Z elements, like Sb. Incorporation of SOC in the calculation splits
the bands and make the band structure more densely packed. However,
this does not remove the contribution from the \emph{s}-states, and
it does not open the band gap. Since the mBJ potential corrects the
conventional LDA \cite{Kim1} type of exchange correlations by
considering  the effect of holes, it creates the proper  separation
between the levels near E$_F$ \cite{Tran1}. The mBJLDA exchange
potential separates the \emph{s}-like states from the
\emph{p}-states near E$_F$ (see Fig. \ref{fat-HI} (c)--(d)), which
were wrongly mixed up at the LDA level. Under this scheme the
Sb-\emph{p} states dominate near $E_F$, as expected \cite{Cardona1}.
When SOC is included over the mBJLDA corrected bands, it provides
the necessary symmetry breaking at the $\Gamma$ point, which in turn
leads to a correct band structure (Fig. \ref{fat-HI} (e)--(f)). We
could apply the above arguments to explain  the role of mBJLDA
exchange potential and SOC in determining correct electronic band
structure of other systems under study.

\begin{figure}[!h]
\begin{center}
\includegraphics[width=0.5\columnwidth]{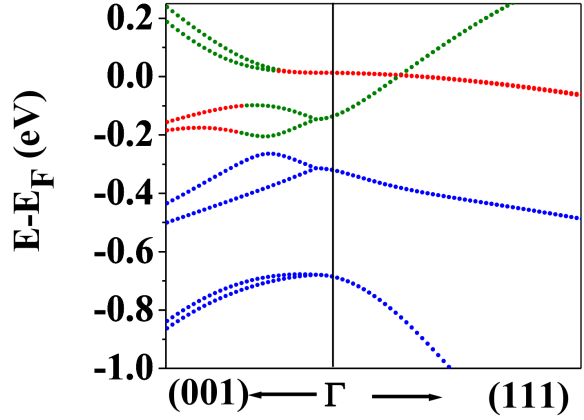}
\caption{The electronic band
 structure of InSb$_{0.5}$As$_{0.5}$ using the mBJLDA method along with
 SOC.} \label{Winkler}
\end{center}
\end{figure}

 The band structure of InSb$_{0.5}$As$_{0.5}$ with CuPt ordering,
calculated using mBJLDA potential, with SOC is shown in Fig.
\ref{Winkler}. We obtain stated \cite{Winkler1} topological features
like band inversion at the $E_F$ level and the existence of the
triple point. A novel topological phase is observed in this system
through appearance of the triple point in the band structure. Our
calculated band structure is an excellent match with that reported
using HSEO6 \cite{Winkler1}. Here also we find that mBJLDA+SOC
scheme is reliable  to obtain details of electronic band structure.

\begin{figure}[!h]
\vskip 0.3cm
\hspace{-0.1in}\includegraphics[scale=.6]{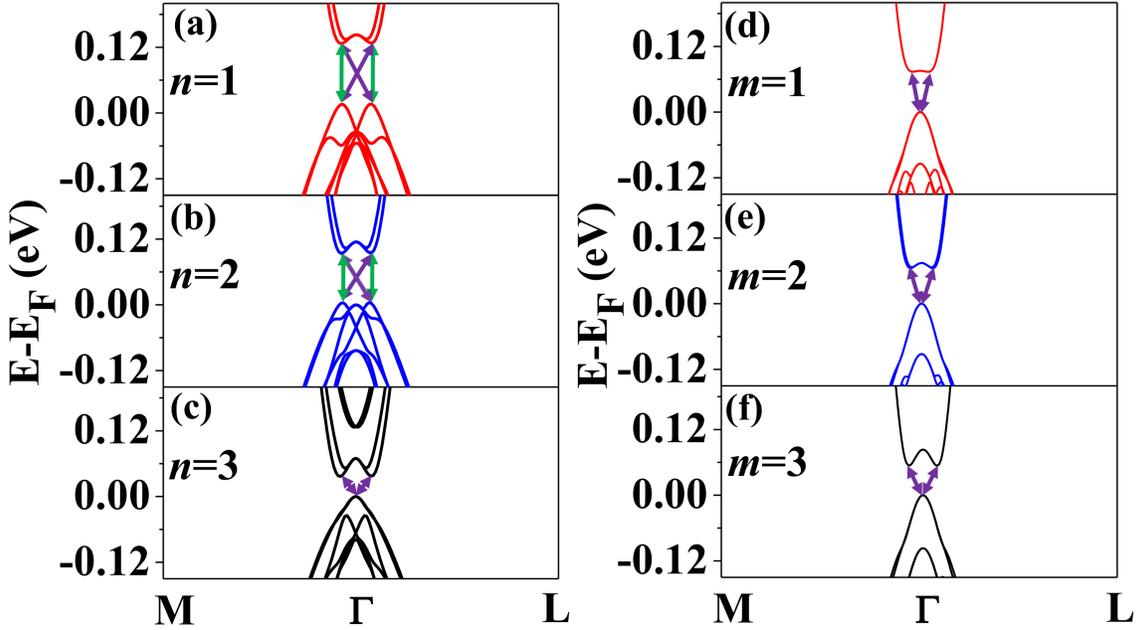}
\caption{\label{f2} Zoomed view of the band structures near $\Gamma$
point for (a)--(c) H-I with \emph{n}=1,2 and 3 (d)--(f) H-II with
\emph{m}=1,2 and 3.} \label{band-layer}
\end{figure}

Next we take a close look at band structures near $\Gamma$ point for
 HSs under study. Refer to Fig. \ref{band-layer} (a)--(c) for H-I
with \emph{n}=1--3. In the figure the direct and indirect band to
band transitions are shown by green and violet lines,
respectively. 
For H-I with \emph{n}=1,2 four transitions, two direct (shown by
green) and two indirect (shown by violet lines) transitions are
possible. All direct and indirect gaps are equal. The obtained
values of the gaps are 0.143 eV and 0.115 eV for H-I with \emph{n}=1
and $n$=2, respectively. However, the band structure gets modified
as the InSb layer thickness is further increased to $n=3$. It has
two indirect band gaps of same energy, 0.54 eV. On the contrary,  no
qualitative change in the nature of the band structure was observed
as we vary the InAs layer thickness in H-II (Fig. \ref{band-layer}
(d)--(f)). The systems, with \emph{m}=1--3, exhibit two indirect
band gaps of equal energy. Table-\ref{bandgap} lists the gap
energies (E$_g$s) of the HSs.
 H-I shows a decrease of the gap from 0.143 to 0.054 eV for
$n=$1--3. However, for H-II with $m$=1--3 the band gap changes only
from 0.075 to 0.084 eV. The variation in band gap is an order of
magnitude higher in H-I than in H-II with increase in the thickness
of the layers.

\begin{table}[ht]
\setlength\belowcaptionskip{5pt}\caption{The bandgap (E$_g$) and number of carriers pockets in H-I and H-II.} 
\centering %
\begin{tabular}{c c c c c c c c c c c c c c c c c c c}
\hline
HS  &&E$_g$ & hole &electron & \\
  && in eV& pockets &pockets &\\
  \hline\hline
   & \emph{n}     \\
&1 &0.143& 6&1&\\
\textbf{H-I} &2 &0.115& 7&6&\\
&3 &0.054& 1&-&\\
\hline
    & \emph{m}     \\
&1 &0.075& 1&-&\\
\textbf{H-II} &2 &0.074& 1&-&\\
&3 &0.084& 1&-&\\
 \hline\hline
\setlength\belowcaptionskip{5pt}
\end{tabular}
\label{bandgap}
\end{table}

Doping or external field is often used to shift the E$_F$ in a
system \cite{Zhu1,Yuan1,Song1}. A close inspection of the band
structures of both HSs near the $\Gamma$ point reveals that a small
shift in E$_F$ is expected to generate different number of Fermi
pockets. A positive effective mass defines an electron pocket while
a hole pocket has a negative effective mass.
The curvature of the bands define the effective mass either to be
positive or to be negative.  Refer to Table-\ref{bandgap}. With a
shift in E$_F$, six hole pockets (three along equivalent
$\Gamma$-$M$ and similarly three along $\Gamma$-$L$ direction) can
be obtained in the band structure of H-I with $n$=1. In addition, we
find the possibility of having an electron pocket only at the
$\Gamma$ point. With increase in the layer thickness, i.e. for $n=$2
the number of hole and electron pockets increased to seven and six
respectively. However, the number of carrier pockets reduced to one
for $n$=3. With a shift in $E_F$, the band structure of H-II with
$m$=1--3 (see Fig. \ref{band-layer} (d)--(f)) is expected to exhibit
only one hole pocket. Thus, we demonstrate that by varying the InSb
layer thickness, one can engineer different carrier pockets (CPs) in
H-I. CPs at a low symmetry points are enhances material
functionality. For example, materials with large numbers of CPs are
ideal for thermoelectric devices \cite{Rabin1,Zhou1}.

\begin{figure}[!h]
\centerline{\epsfxsize=6.5in\epsffile{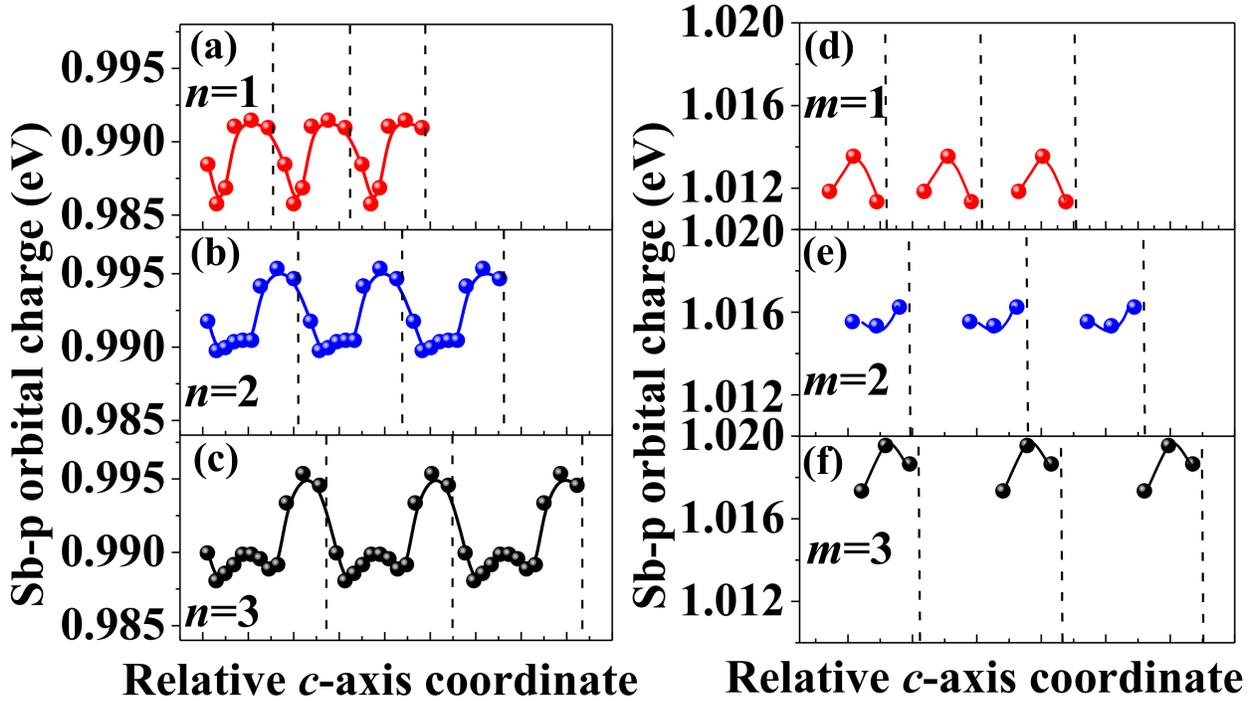}}
\caption{Charge distribution of Sb-p orbital with relative c -axis
coordinate for (a)--(c) H-I and (d)-(f) H-II. Dotted lines mark the
boundary of one unit cell of the HSs. \label{OS}}
\end{figure}

In above we observe that the intricate features of electronic band
structures of H-I and H-II are markedly different. To find the
possible origin of this difference in evolution characteristic of
the electronic band structures of H-I and H-II with layer thickness,
we looked into the charge modulation along the $c$ axis of the  HSs.
As seen through the fat band analysis,  Sb \emph{p}-orbitals of
these systems have a dominant contribution near $E_F$. We studied
the variation of charge on Sb sites along the $c$ axis, as shown in
Fig. \ref{OS} (a)--(c) and (d)--(f) for H-I and H-II respectively.
In H-I, as we change the InSb layer thickness, the number of Sb
atoms increases from six to twelve and they are inequivalent (see
Fig. \ref{OS} (a)-(c)). Thus, the increase in InSb layer thickness
in H-I has two consequences. First it shrinks the BZ along the
Z-direction and secondly, it increases the number of states within
the BZ (due to increase in inequivalent Sb atoms). As a consequence,
the band structure significantly changes when $n$ changes from 1 to
3. In H-II, with an increase in the InAs layer thickness, the BZ
shrinks. However,  the numbers of relevant inequivalent atoms remain
three for $m$=1--3 (see  Fig. \ref{OS} (d)--(f)). Hence, we do not
find appreciable change in the band structure for this system.

\section{Conclusion}
In summary, we have calculated electronic band structure of two
different HSs, one component of which is a topological semimetal of
CuPt ordered InSb$_{0.5}$As$_{0.5}$ and other component is a band
insulator. The HSs are also semiconductors. We have shown that for
InSb/InAs$_{0.5}$Sb$_{0.5}$  HS, the band gap as well as fine
features near the Fermi level can be modulated by varying the
thickness of the InSb segment.  Another important take away from
this work is how differently these HSs would respond to a small
perturbation owing to the difference in the number of their Fermi
pockets.

\noindent\textbf{Acknowledgements}\\
Authors thank Professor Debraj Choudhury, IIT Kharagpur, for
valuable discussion. Authors also acknowledge the use of the
computing facility from the DST-Fund for Improvement of S\&T
infrastructure ( phase-II) Project installed in the Department of
Physics, IIT Kharagpur, India. AR thanks Department of Science and
Technology, Government of India, for financial assistance.


\end{document}